\documentclass[aps,preprintnumbers,
amsmath,amssymb,prb,superscriptaddress,longbibliography
, groupedaddress,twocolumn
]{revtex4-2}
\UseRawInputEncoding
\usepackage{dcolumn}% Align table columns on decimal point
\usepackage{color}

\usepackage[caption=false]{subfig}
\usepackage{feynmp}
\usepackage{feynmp-auto}
\usepackage{float}
\usepackage{bbm}
\usepackage{empheq}
\usepackage{appendix}
\def\comment#1{}

\newcommand{\nc}{\newcommand}
\nc{\beq}{\begin{eqnarray}}
	\nc{\eeq}{\end{eqnarray}}
\nc{\scs}{\scriptstyle}
\nc{\setval}{\fmfset{wiggly_len}{3mm} \fmfset{arrow_len}{1.5mm}
	\fmfset{arrow_ang}{13} \fmfset{dash_len}{1.5mm}\fmfpen{0.125mm}
	\fmfset{dot_size}{2thick}}

\usepackage{bm,latexsym,mathrsfs,enumerate,color}
\usepackage[mathcal]{euscript}
\usepackage[breaklinks=true,unicode=true,urlcolor = blue,colorlinks = true,citecolor = blue,linkcolor = blue]{hyperref}
\usepackage{graphicx}
\usepackage{todonotes}
\usepackage{wrapfig}

\renewcommand{\vec}[1]{\bm{#1}}

\def\slashchar#1{\setbox0=\hbox{$#1$}           % set a box for #1
	\dimen0=\wd0                                 % and get its size
	\setbox1=\hbox{/} \dimen1=\wd1               % get size of /
	\ifdim\dimen0>\dimen1                        % #1 is bigger
	\rlap{\hbox to \dimen0{\hfil/\hfil}}      % so center / in box
	#1                                        % and print #1
	\else                                        % / is bigger
	\rlap{\hbox to \dimen1{\hfil$#1$\hfil}}   % so center #1
	/                                         % and print /
	\fi}                                         %

\DeclareMathAlphabet\mathbfcal{OMS}{cmsy}{b}{n}

\usepackage{physics}
\usepackage{xcolor}

\usepackage{physics}

\begin{document}
%\onecolumngrid
\title{Skyrmion-vortex pairing and vortex-drag induced Skyrmion Hall effect}
\author{Shantonu Mukherjee}
\email{shantanumukherjeephy@gmail.com}
\affiliation{Indian Institute of Technology, Bombay, India}
\begin{abstract}
    An interaction between ferromagnetic and superconducting orders, to be realized in a two dimensional ferromagnetic superconductor, is proposed obeying necessary symmetry principles. This interaction allows us to formulate a duality, similar to the Boson-vortex duality in 2+1 dimensional superfluid. In the dual theory the Skyrmion and the vortex excitations interact with each other via an emergent gauge field. The static interaction potential is attractive for a Skyrmion and a vortex with opposite topological charges. This interaction can lead to formation of bound pairs of the mentioned topological excitations. Furthermore, we argue that such pairing implies that a Magnus force acting on the vortex induces a transverse, Hall-like drift motion of the Skyrmion, which we term the vortex-drag induced Skyrmion Hall effect. Possible experimental manifestations of this effect are also discussed. 
\end{abstract}
\date{\today}
\maketitle
\section{Introduction}

The coexistence of ferromagnetism and superconductivity, though seemingly counterintuitive, has been experimentally realized in various material systems owing to recent advances in experimental techniques~\cite{Oner_2020, PhysRevB.74.094518, https://doi.org/10.1002/pssa.2210990104, Zhu2016SignatureOC}. This coexistence phase provides a fertile ground for exploring rich physical phenomena emerging from the interplay between topological excitations such as vortices in the superconducting order and magnetic Skyrmions. Among the most intriguing directions of current research is the realization of hybrid or composite topological entities formed by the coupling of superconducting vortices and magnetic Skyrmions at superconductor--ferromagnet interfaces\cite{PhysRevB.99.014511, PhysRevLett.122.097001, PhysRevLett.117.017001, PhysRevB.103.174519, PhysRevB.102.014503, PhysRevLett.126.117205} as well as within ferromagnetic superconducting bulks~\cite{Mukherjee:2024tih, leask2026interactions}. In these studies, the internal structure of the composite objects and their mutual interactions have been analyzed through collective field descriptions, often by solving the corresponding Euler-Lagrange equations. While that approach captures the microscopic details faithfully, it often obscures an intuitive understanding of the effective macroscopic dynamics of these collective excitations.\\

To gain such physical intuition, it is convenient to adopt an effective point-particle like description of these mesoscopic topological entities. In this framework, the dynamics of each composite object can be tracked via the motion of its center, enabling one to employ the language of classical particle mechanics. This description allows a natural characterization of local interactions between these emergent particles. One systematic way to construct such an effective point-particle description is through duality transformations that map collective field configurations to an effective theory of point particles with emergent local interactions between these particle like local degrees of freedom.\\

A well-known example of this correspondence is the particle-vortex duality in (2+1)-dimensional field theories~\cite{Peskin:1977kp,PhysRevLett.47.1556}. This duality plays a central role in analyzing the Berezinskii-Kosterlitz-Thouless (BKT) transition~\cite{1973JPhC....6.1181K} and is typically discussed in the context of the planar X–Y model--an effective low-energy description for two-dimensional superfluids or spin systems~\cite{RevModPhys.52.453}. The particle-vortex duality maps vortex excitations into point particles interacting via an effective logarithmic potential. Such interactions lead to bound vortex-antivortex pairs forming a confined phase, while the BKT transition describes a transition to a deconfined phase in which isolated vortices emerge. Moreover, in the dual formulation, the vortex dynamics is characterized in terms of the motion of its core, which provides a powerful tool to study vortex motion and response under external perturbations such as electromagnetic fields. \emph{Constructing an analogous duality in the present context would yield an equally powerful framework for analyzing the dynamics of Skyrmion-vortex bound states}.\\

In this work, we begin with a theoretical model akin to that introduced in a preceding study~\cite{Mukherjee:2024tih} by the author. The model consists of a nonlinear sigma model coupled to the Landau–-Ginzburg theory of superconductivity through a spin--magnetic-field interaction (SMFI) term, augmented by an additional direct coupling term (discussed in Section~\ref{sec-iii}). The inclusion of this direct coupling term is pivotal for formulating a dual description to be presented in this article. As will be demonstrated, the dual theory features an emergent gauge field mediating an interaction between the Skyrmion and the vortex excitations. Depending on the field configuration, this interaction can be attractive for the Skyrmion-antivortex or anti-Skyrmion-vortex pairs, leading to the formation of bound states. Notably, this direct coupling, essential for realizing the dual mapping, has not been explored previously in the context of magnetic systems to our knowledge, making the duality developed here novel. (For comparison, see \cite{PhysRevLett.64.1313}~for an analogous duality in fermionic systems.)\\
\begin{figure}
    \centering
    \includegraphics[clip, trim=1cm 15cm 0.2cm 7cm, width=1\columnwidth]{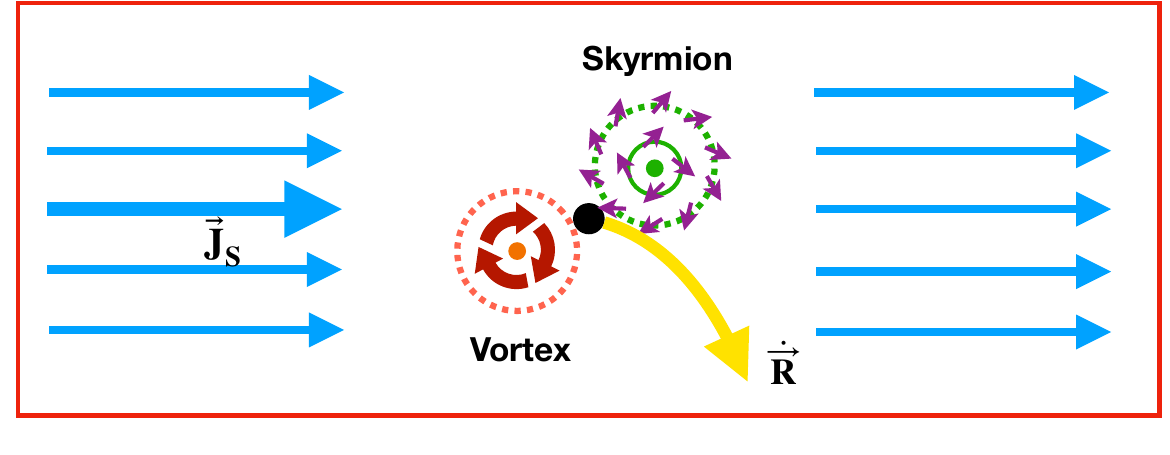}
    \caption{This is a schematic depiction of vortex-drag induced Skyrmion Hall drift. The blue arrows show the direction of the supercurrent. The black dot in the midst of the Skyrmion and the vortex represents the centre of mass of the composite. The yellow arrow shows the motion of the centre of mass of the composite perpendicular to the supercurrent.}
    \label{fig:placeholder}
\end{figure}

The dual theory provides a natural framework to study the coupled dynamics of the Skyrmion-vortex bound states via coupled Thiele-type equations. Assuming rigid adiabatic dynamics of the collective topological objects in which their motion does not lead to significant deformation of their internal structures, we shall derive the Thiele-like equation of motion~\cite{PhysRevLett.30.230, PhysRev.140.A1197, 1966PMag...14..667N} of their centres denoted by $\vec{R}_v$ and $\vec{R}_s$ for the vortex and for the Skyrmion respectively. Due to the Skyrmion-vortex interaction, these equations are coupled. Decoupling these equations in terms of centre of mass coordinate and relative coordinate, and under certain approximations, we shall argue that a Magnus force on the vortex will create a Hall-like drift motion of the centre of mass, leading to a vortex-drag induced Skyrmion Hall effect. We have also discussed possible experimentally detectable consequences of such a novel effect. The dual theory also offers a theoretical avenue to explore BKT-like unbinding transitions of these composite excitations--an exciting new direction in the field.\\

The paper is organized as follows. In Sec.~\ref{sec-ii} we review the $\mathbb{C}\text{P}^1$ representation of the spin system. In Sec.~\ref{sec-iii} we present the model of a ferromagnetic superconductor in two dimensions and derive a dual description of this model utilizing the $\mathbb{C}\text{P}^1$ representation of spin. In Sec.~\ref{sec-iv} we use the dual theory to derive coupled equations of motion of vortex and Skyrmion and show how, under certain approximations, a vortex-drag induced Skyrmion Hall effect arises. We end our discussion by laying out possible experimental implications of our results and also discuss some interesting future directions in Sec.~\ref{sec-v}.

\section{Revisiting the $\mathbb{C}\text{P}^1$ representation of spin system}\label{sec-ii}
Two-component complex spinors form a two dimensional complex Hilbert space $\mathbb{C}^2$. However, for any given state $\ket{\psi}$ there will be states $\ket{\psi^\prime}=\beta \ket{\psi}$ which will differ by a complex phase factor $\beta\in \mathbb{C}^*$. These states form an equivalent class of states. Thus, the Hilbert space is a collection of such equivalent classes or rays. The space formed by these physical states or rays is called a projective Hilbert space or $\mathbb{C}\text{P}^1$ (Quotient space $(\mathbb{C}^2 \setminus\{0\})/ \mathbb{C}^*$). For Normalized spinors $|z|^2=1$, the $\mathbb{C}\text{P}^1$ space becomes the Bloch sphere. This is because $|z|^2=1$ defines a 3-sphere, while the space of equivalent states connected by $U(1)$ transformations forms a circle leading to a quotient space $S^3/ U(1)$ which is topologically equivalent to $S^2$. Thus, the space of normalized 2d rays forms a 2-sphere known as Bloch sphere.\\

The fluctuation in a spin system is expressed in terms of the normalized spin vector $\hat{n}$. The spin states expressed by these vectors lie on a 2-sphere, which is again the Bloch sphere. This induces a map between normalized spin vector $\hat{n}$ and $\mathbb{C}\text{P}^1$ spinor expressed by the following relation
\begin{equation}\label{cp1}
    \hat{n}= z^\dagger \vec{\sigma} z,\,\,\,\, z^\dagger z=1.
\end{equation}
The 2d spinor $z$ can be chosen as 
\begin{equation}
    z= \begin{pmatrix}
        \cos (\theta/2)\\
        e^{i\phi} sin (\theta/2)
    \end{pmatrix}
\end{equation}
Let us now see how various terms in the action of a ferromagnetic system change in this new representation. We first write down the general action for the unit spin field for such systems
\begin{equation}\label{spin_action_FM}
    S_{FM}= \int d^3x \left(\vec{\mathcal{A}}(\hat{n})\cdot \partial_t \hat{n} - \frac{\rho_M}{2}\left(\mathbf{\grad} n^i\right)^2 - \mu n^i \varepsilon^{ijk} \partial^j A^k\right)
\end{equation}
Here, the first term corresponds to the Berry phase associated with the time evolution of the spin, arising from the solid angle traced by the spin configuration on the Bloch sphere. This term arises in the following way 
\begin{eqnarray}\label{spin_Berry}
    L = i \bra{\hat{n}}\partial_t\ket{\hat{n}} - H = i \bra{\hat{n}}\partial_{n^i}\ket{\hat{n}} \frac{\partial n^i}{\partial t} - H 
\end{eqnarray}
One can recognize that the first term in the second line of Eq~\eqref{spin_Berry} is the Berry connection $\vec{\mathcal{A}}=i \bra{\hat{n}}\grad_{\hat{n}}\ket{\hat{n}} $. Presence of this geometric phase in a dynamical spin action can be rigorously derived using spin coherent states~\cite{Han:2017fyd}. It can be readily verified that in the new $\mathbb{C}\text{P}^1$ representation, this geometric phase term takes the following form
\begin{equation}
    \vec{\mathcal{A}}\cdot \partial_t \hat{n}=i \bra{\hat{n}}\grad_{\hat{n}}\ket{\hat{n}}\cdot \partial_t \hat{n}= S (\cos\theta -1)\Dot{\phi}= i z^\dagger \partial_t z,
\end{equation}
where the expression in the third step is obtained using spin coherent states for $\ket{\hat{n}}$~\cite{Han:2017fyd}. Now we take a look at the second term of Eq~\eqref{spin_action_FM}. This term is called a non-linear sigma term which captures the fluctuation of the spin field $\hat{n}$. The constant $\rho_M$ is the spin stiffness, which determines the energy cost associated with spatial variations of the spin configuration. In the new $\mathbb{C}\text{P}^1$ representation this term will have the following form
\begin{equation}\label{nlsm_cp1}
    \frac{\rho_M}{2}\left(\mathbf{\grad} n^i\right)^2 = 2 \rho_M \left((\grad z^\dagger)\cdot(\grad z) - (z^\dagger\grad z)\cdot (\grad z^\dagger z)\right),
\end{equation}
where we have used the identity 
\begin{equation}
    \sigma^i_{a b} \sigma^i_{\alpha \beta}= 2 \delta_{a\beta} \delta_{b\alpha} - \delta_{ab}\delta_{\alpha \beta}
\end{equation}
and $z^\dagger z=1$. The term $z^\dagger\grad z$ or more generally $a_{\mu}=-iz^\dagger\partial_\mu z$ behaves as a gauge field under the transformation $z\rightarrow e^{i\chi} z$:
\begin{equation}
    z\rightarrow e^{i\chi} z,\quad a_{\mu} \rightarrow a_{\mu} + \partial_\mu \chi.
\end{equation}
Due to this gauge degree of freedom the Eq~\eqref{nlsm_cp1} can be written by defining a covariant derivative for $z$ as
\begin{equation}
    \frac{\rho_M}{2}\left(\mathbf{\grad} n^i\right)^2= 2 \rho_M |D_i z|^2,\quad D_i=\left(\partial_i - i a_i\right).
\end{equation}
Thus, we see that $\mathbb{C}\text{P}^1$ representation brings out a gauge degree of freedom in the effective theory. This will help us further in thinking of new terms that can be present in the theory, as we shall see in the next section. In this context, let us also discuss one other term in this representation that will be useful in the formulation of the duality. This term is the topological current density or the Skyrmion current density and is expressed as
\begin{equation}\label{Skyrmion_current}
    J^s_\alpha = \frac{1}{2} \varepsilon_{\alpha\beta\gamma} \varepsilon^{i j k} n^i \partial^\beta n^j \partial^\gamma n^k,
\end{equation}
where $i,\,j,\,k \in (1,2,3),\quad \alpha,\,\beta,\, \gamma\in (0,1,2)$. Using the relation in Eq~\eqref{cp1} into the above definition, one can show that the Skyrmion current density $J^s_\mu$ takes the following form
\begin{equation}
    J^s_\alpha= \varepsilon_{\alpha\beta\gamma} \partial^\beta a^\gamma.
\end{equation}
This form allows us to think of the Skyrmion current as an effective magnetic field ($\vec{\tilde{B}}$) and effective electric field ($\vec{\tilde{E}}$)
\begin{align}
    J^s&= (\partial_x a_y - \partial_y a_x,\,\, \partial_y a_t - \partial_t a_y,\,\, \partial_t a_x - \partial_x a_t)\nonumber\\ &=  (\tilde{B}_z,\,\, -\tilde{E}_y,\,\, \tilde{E}_x).
\end{align}
Due to this form, the conservation law of this topological current transforms to Faraday's Law 
\begin{equation}
    \partial_\mu J^{s\mu} = -\partial_t \tilde{B}^z + (-\partial_x \tilde{E}_y+ \partial_y \tilde{E}_x)=0.
\end{equation}

\section{Skyrmion-vortex Pairing from duality}\label{sec-iii}
In this work, our system of choice is a two dimensional ferromagnetic superconductor thin film hosting both s-wave superconducting and ferromagnetic order, which coexist and interact with each other. An effective theory for such a system can be expressed by the following Lagrangian
\small{\begin{align}
& \mathcal{L}_T= \mathcal{L}_{SC} + \mathcal{L}_{FM} + \mathcal{L}_I,~\label{total_lagrangian}\\ &
\mathcal{L}_{SC}= \rho_s(\partial_t\chi + i \eta_v^*\partial_t \eta_v + q A_0)-\frac{\bar{\rho}_s}{2m}(\mathbf{\grad}\chi + i\eta_v^*\mathbf{\grad}\eta_v + q\vec{A})^2\nonumber\\ & - \frac{g}{2} \left(\rho_s - \bar{\rho}_s\right)^2.\\ &
\mathcal{L}_{FM}= \vec{\mathcal{A}}(\hat{n})\cdot \partial_t \hat{n} - \frac{\rho_M}{2}\left(\mathbf{\grad} n^i\right)^2 - \mu n^i \varepsilon^{ijk} \partial^j A^k.\\ &
\mathcal{L}_I= - \frac{\lambda}{2}(1-\cos\theta)\mathbf{\grad}\phi \cdot \sqrt{\bar{\rho}_s}(\mathbf{\grad}\chi + i\eta_v^*\mathbf{\grad}\eta_v + q\vec{A}).~\label{proposed interaction}
\end{align}}

%%%%%%%%%%%%%%%%%%%%%%%%%%%%%%%%%%%%%%%%%%%%%%%%%%%%%%%%%%%%%%%%%%%
%%%%%%%%%%%%%%%%%%%%%%%%%%%%%%%%%%%%%%%%%%%%%%%%%%%%%%%%%%%%%%%%%%%
The first part of the total Lagrangian expresses the phase fluctuation of the superconducting order parameter~\cite{Lee:1991jt} in the symmetry broken ground state where the amplitude part of the order parameter is a constant in space-time: $\psi= \sqrt{\rho_s}e^{-i\chi} \eta_v$. Here, the field $\chi$ describes the fluctuating phase of the Cooper pair wave function and $\eta_v= e^{-i\chi_v}$ is the singular part of the phase, which takes into account the vortex configurations. Also, here $\bar{\rho}_s$ represent the mean field density of Cooper pairs in the ground state. This action for the superconducting phase fluctuation emerges from the Gross-Pitaevskii action for the order parameter $\psi$. The second term in Eq~\eqref{total_lagrangian} expresses the spin fluctuation of the ferromagnet~\cite{PhysRevLett.75.3509}. The angles $\theta,\phi$ define the components of the unit spin field $\hat{n}\, (\sin\theta \cos\phi,\, \sin\theta \sin\phi,\, \cos\theta)$ which describes the spin coherent states on the Bloch sphere in the spin space. We have also written the spin - Magnetic field interaction (SMFI) in this part of the Lagrangian for the sake of generality. However, in the following discussion, we shall ignore this term. The last term $\mathcal{L}_I$ expresses an interaction between the two coexisting phases. This new interaction should be invariant under infinitesimal global SO(3) rotation of the spin field $\hat{n}$ as it is the symmetry of the system considered here. {It is shown in Appendix-\ref{appendix-A} that, although the Lagrangian density is not strictly invariant under global SO($3$), the action is invariant upto a boundary term. It follows from the fact that the term $\tfrac{1}{2}(1-\cos\theta)\mathbf{\grad}\phi$ is an explicit form of the emergent gauge field $a^i$ introduced in Sec-\ref{sec-ii} and that the coupled term $(\mathbf{\grad}\chi + i\eta_v^*\mathbf{\grad}\eta_v + q\vec{A})$ is proportional to the conserved current of the superconducting component.}

In the following, we shall adopt the $\mathbb{C}\text{P}^1$ representation of the spin field and shall formulate a duality which will lead to a short range attractive interaction of vortex and Skyrmion with opposite topological charges. Using the $\mathbb{C}\text{P}^1$ representation of various relevant term developed in Sec-\ref{sec-ii} the Lagrangian in Eq~\eqref{total_lagrangian} takes the following form
\begin{widetext}
   \begin{align}\label{spinon rep}
 \mathcal{L}= &   \rho_s\left(\partial_t\chi + i \eta_v^*\partial_t \eta_v + q A_0 - \frac{1}{\rho_s} a_0\right) - \frac{\bar{\rho}_s}{2m}\left(\mathbf{\grad}\chi + i\eta_v^*\mathbf{\grad}\eta_v + q\vec{A}\right)^2\nonumber \\ &- \frac{g}{2} \left(\rho_s - \bar{\rho}_s\right)^2  - \frac{\rho_M}{2} 4(|D_i z|^2) - \lambda \sqrt{\bar{\rho}_s}\vec{a} \cdot (\mathbf{\grad}\chi + i\eta_v^*\mathbf{\grad}\eta_v + q\vec{A}),
\end{align} 
\end{widetext}
 We note that the emergent gauge field $a_\mu$ has appeared in the first, fourth, and the last term. In the original form, the theory had the symmetry under a global SO$(3)$ transformation of the spin field $\hat{n}$ which in this new representation translates to the invariance of the Lagrangian under global $SU(2)$ transformation of the spinor field $z$. Also, due to the appearance of the gauge field $a_\mu$ a new gauge invariance under the transformation $\mathbf{a}\rightarrow \mathbf{a} + \mathbf{\grad}\xi $ emerges. This is manifestly true for the fourth term in Eq~\eqref{spinon rep}, where as in the last term this invariance is true as the associated current $\mathbf{J}= \left(\mathbf{\grad}\chi + i\eta_v^*\mathbf{\grad}\eta_v + q\vec{A}\right)$ is the conserved current for the superconductor. This gauge invariance of the last term suggests that one can include this interaction term in this $\mathbb{C}\text{P}^1$ representation purely on the basis of gauge symmetry. We shall now derive the theory dual to the Lagrangian in Eq~\eqref{spinon rep} following the derivation of Boson-vortex duality or particle-vortex duality in 2+1d~\cite{Lee:1991jt, PhysRevLett.64.1313, Mukherjee:2019vmi, Franz:2006gb, PhysRevB.39.2756, PhysRevLett.47.1556}. In this procedure, we shall first combine the second and the last term in Eq~\eqref{spinon rep} by completion of square. Subsequently, we shall linearize the squared term resulting from the earlier procedure by introducing a Hubbard-Stratonovich decoupling field. Integrating out the small scale fluctuation in the phase field $\chi$ in the next step will lead to an emergent gauge field $b_\mu$ whose curl is dual to the superfluid velocity $\partial_\mu\chi$. Detailed steps of this dualization procedure are given in the Appendix~\ref{appendix-B}. This will give us the following dual Lagrangian.
 %%%%%%%%%%%%%%%%%%%%%%%%%%%%%%%%%%%%%%%%%%%%%%%%%%%%%%%%%%%%
 %%%%%%%%%%%%%%%%%%%%%%%%%%%%%%%%%%%%%%%%%%%%%%%%%%%%%%%%%%%%
 %%%%%%%%%%%%%%%%%%%%%%%%%%%%%%%%%%%%%%%%%%%%%%%%%%%%%%%%%%%%
 %%%%%%%%%%%%%%%%%%%%%%%%%%%%%%%%%%%%%%%%%%%%%%%%%%%%%%%%%%%%%%
 %%%%%%%%%%%%%%%%%%%%%%%%%%%%%%%%%%%%%%%%%%%%%%%%%%%%%%%%%%%%%%
 %%%%%%%%%%%%%%%%%%%%%%%%%%%%%%%%%%%%%%%%%%%%%%%%%%%%%%%%%%%%%%

\begin{widetext}
      \begin{align}\label{dual_lagrangian_1}
   \mathcal{L}= \frac{m}{8\pi^2\bar{\rho}_s} (\partial_0 b_i-\partial_i b_0 )^2  - \frac{g}{8\pi^2} \left(\grad \times \vec{b} - 2\pi \bar{\rho}_s\right)^2+ \frac{1}{2\pi} \epsilon_{\mu\nu\rho} \partial^\nu b^\rho\left( i \eta_v^*\partial^\mu \eta_v + q A^\mu + \tilde{a}^\mu\right) - 2 \tilde{\rho}_M (|\mathbf{\grad} w|^2 - |\vec{a}^\prime|^2).
   \end{align}
 \end{widetext}

%%%%%%%
The Lagrangian derived above in Eq~\eqref{dual_lagrangian_1} is the dual theory containing an interaction between vortex and Skyrmion via the emergent gauge field $b_\mu$. This can be realized by taking a closer look at the third term of the Eq~\eqref{dual_lagrangian_1}.
After an integration by parts and ignoring the boundary terms, one can show that the gauge field $b_\mu$ couples to both the vortex current $J^\mu_v$ and the Skyrmion current $J^\mu_s$ defined in the following way
\begin{eqnarray}
    J^v_\mu=  \epsilon_{\mu\nu\rho} \partial^\nu (i \eta_v^*\partial^\rho \eta_v) ,\,\,\, J^s_\mu= \epsilon_{\mu\nu\rho} \partial^\nu a^\rho. 
\end{eqnarray}
and thus the vortex and Skyrmion in this theory interact via this emergent gauge field $b_\mu$. The nature of this interaction can be understood more clearly if one determines the static interaction potential between the vortex and the Skyrmion charge density, which can be obtained by integrating over $b_\mu$. To proceed, we first note from Eq~\eqref{dual_lagrangian_1} that the emergent gauge field $b_\mu$ couples to the dynamical electromagnetic gauge field $A_\mu$ forming a mixed Chern-Simons or a ``BF" term in 2+1d which leads to the generation of mass of $b_\mu$ when $A_\mu$ is integrated out~\cite{Deser1982TopologicallyMassive, Allen1991TopologicalMass, Mukherjee:2019vmi}. Thus to integrate out $b_\mu$, one first needs to obtain a ``diagonalized'' action of $b_\mu$ by integrating over $A_\mu$ removing the off-diagonal ``BF'' coupling term.  In the following, we shall integrate out the gauge fields $A_\mu$ and subsequently $b_\mu$.
%%%%%%%%%%%%%%%%%%%%%%%%%%%%%%%%%%%%%%%%%%%%%%%%%%%%%%%%%%%%%%%%%%%%%
%%%%%%%%%%%%%%%%%%%%%%%%%%%%%%%%%%%%%%%%%%%%%%%%%%%%%%%%%%%%%%%%%%%%%
%%%%%%%%%%%%%%%%%%%%%%%%%%%%%%%%%%%%%%%%%%%%%%%%%%%%%%%%%%%%%%%%%%%%%

%%%%%%%%%%%%%%%%%%%%%%%%%%%%%%%%%%%%%%%%%%%%%%%%%%%%%%%%%%%%%%%%%%%%%%%%%%
%%%%%%%%%%%%%%%%%%%%%%%%%%%%%%%%%%%%%%%%%%%%%%%%%%%%%%%%%%%%%%%%%%%%%%%%%%
%%%%%%%%%%%%%%%%%%%%%%%%%%%%%%%%%%%%%%%%%%%%%%%%%%%%%%%%%%%%%%%%%%%%%%%%%%
The detailed steps of integrating out the electromagnetic gauge field $A_\mu$ are given in Appendix~\ref{appendix-C}. After integrating out the vector field $A_\mu$, we get the following new dual Lagrangian 
\begin{widetext}
    \begin{align}
    \mathcal{L}= &\frac{m}{8\pi^2\bar{\rho}_s} (\partial_0 b_i-\partial_i b_0 )^2  - \frac{g}{8\pi^2} \left(\grad \times \vec{b} - 2\pi \bar{\rho}_s\right)^2+ \frac{1}{2\pi} \epsilon_{\mu\nu\rho} \partial^\nu b^\rho\left( i \eta_v^*\partial^\mu \eta_v + \tilde{a}^\mu\right)\\ & - \frac{q^2}{16\pi^2} (\varepsilon^{\mu\nu\lambda}\partial_\nu b_\lambda) (\square)^{-1} (\varepsilon_{\mu\alpha\beta}\partial^\alpha b^\beta)  - 2 \tilde{\rho}_M (|\mathbf{\grad} w|^2 - |\vec{a}^\prime|^2).
\end{align}
\end{widetext}
Obtaining the full action of $b_\mu$, the next step for obtaining the interaction potential between a vortex and a Skyrmion is to integrate out $b_\mu$. To do that, we first notice from Eq~\eqref{auxiliary lagrangian} that the third term in this action imposes a mean field constraint on the Cooper pair density: $J_0 = \frac{1}{2\pi} \grad \times \vec{b} = \bar{\rho}_s$. This leads to the mean field configuration of $\vec{b}$ to be
\begin{equation}
    \vec{b} = \pi \bar{\rho}_s (-y, x).
\end{equation}
Here we note that, although one can consider fluctuations of the field $\vec{b}$ arising from density fluctuation around the mean field value $\bar{\rho}_s$, in this discussion we shall continue with this mean field approximation, neglecting the fluctuations. However, this does not fix the field $b_0$, and it remains to be a dynamical field in the theory. As we shall see below, this dynamical field will lead to an attractive interaction potential between vortex and Skyrmion charges. 
%%%%%%%%%%%%%%%%%%%%%%%%%%%%%%%%%%%%%%%%%%%%%%%%%%%%%%%%%
%%%%%%%%%%%%%%%%%%%%%%%%%%%%%%%%%%%%%%%%%%%%%%%%%%%%%%%%%
%%%%%%%%%%%%%%%%%%%%%%%%%%%%%%%%%%%%%%%%%%%%%%%%%%%%%%%%%

%%%%%%%%%%%%%%%%%%%%%%%%%%%%%%%%%%%%%%%%%%%%%%%%%%%%%%%%%%%
%%%%%%%%%%%%%%%%%%%%%%%%%%%%%%%%%%%%%%%%%%%%%%%%%%%%%%%%%%%
%%%%%%%%%%%%%%%%%%%%%%%%%%%%%%%%%%%%%%%%%%%%%%%%%%%%%%%%%%%
Integrating out $b_0$ (See Appendix~\ref{appendix-D} for details) and putting the Skyrmion and vortex charge density $J_s^0$ and $J_v^0$ as
\begin{equation}
    J_s^0= Q_s \delta^2(\vec{x}- \vec{R}_s),\quad J_v^0= Q_v \delta^2(\vec{x}- \vec{R}_v).
\end{equation}
we get 
%\begin{widetext}
    \begin{align}\label{interaction potential}
    &S_{\text{Int}}=-\int\, dt\, 2 \lambda \sqrt{\bar{\rho}_s} Q_v Q_s   V(|\vec{R}_v(t)- \vec{R}_s(t)|), \\ & V(|\vec{R}_v- \vec{R}_s|)= \int \,\frac{d^2k}{(2\pi)^2} \, \frac{e^{i\vec{k}\cdot (\vec{R}_v- \vec{R}_s)}}{k^2 + \frac{q^2\bar{\rho}_s}{2m} }.
\end{align}
%\end{widetext}
The Eq~\eqref{interaction potential} suggests that the interaction potential between a static Skyrmion and a vortex is given by
\begin{equation}\label{interaction potential form}
    V(|\vec{R}_v- \vec{R}_s|)= \frac{1}{2\pi}K_0(\tilde{m}|\vec{R}_v- \vec{R}_s|),\quad \tilde{m}= \frac{q^2\bar{\rho}_s}{2m}.
\end{equation}
This interaction potential between a Skyrmion and a vortex can be attractive depending on the sign of $\lambda,\, Q_s,\, Q_v$. In particular, for $\lambda>0$ the potential will be attractive for an anti-Skyrmion and a vortex or vice versa. This attractive interaction can lead to the formation of an anti-Skyrmion-vortex pair or vice versa. 
%%%%%%%%%%%%%%%%%%%%%%%%%%%%%%%%%%%%%%%%%%%%%%%%%%%%%%%%%%%%%%%%%
%%%%%%%%%%%%%%%%%%%%%%%%%%%%%%%%%%%%%%%%%%%%%%%%%%%%%%%%%%%%%%%%%
\section{Coupled Dynamics of Skyrmion-vortex composite:}\label{sec-iv}
%%%%%%%%%%%%%%%%%%%%%%%%%%%%%%%%%%%%%%%%%%%%%%%%%%%%%%%%%%%%%%%%%
In this section we shall consider the dual theory derived in the previous section and develop a collective coordinate description of the dynamics of the vortex and the Skyrmion taking into account the attractive potential derived previously. Let us first try to understand the underlying assumptions of such a phenomenological collective dynamics of these topological objects. We focus on the regime where both the vortex and the Skyrmions are well formed localized topological textures whose structure remains unaltered due to their motion. This will happen if the characteristic frequency of their translational motion is small compared to the characteristic frequency of the internal excitations like magnons. In this limit, their dynamics is solely determined by the motion of their centres, denoted by $\vec{R}_v(t)$ and $\vec{R}_s(t)$ respectively.\\

We also remind ourselves that we are considering the emergent gauge field within the mean field approximation, neglecting its fluctuations, and hence it has the configuration $\vec{b}= \pi \bar{\rho}_s (-y, x)$. Using this into our dual Lagrangian of Eq~\eqref{dual_lagrangian_1} we derive the equation of motion of a vortex and a Skyrmion present in the system in the following way. \\

The term relevant to the vortex motion is given by
\begin{equation}\label{eom vortex 1}
    S_v = \int d^3x\, \frac{1}{2\pi} \vec{b}\cdot \vec{J}_v
\end{equation}
Here $\vec{J}_v$ is the vortex current, and considering the motion of the centre of the vortex, this current can have the following form
\begin{equation}
    \vec{J}_v= Q_v (-\dot{X}_v, -\dot{Y}_v )  \delta^2(\vec{x}- \vec{R}_v)
\end{equation}
Putting these expressions into Eq~\eqref{eom vortex 1}, we obtain the full action of the vortex
    \begin{align}\label{eom vortex 2}
   S_v = &\int dt\, \bigg( \frac{1}{2} M_v \vec{\dot{R}_v}^2 + \frac{\bar{\rho}_s Q_v}{2} (\vec{\dot{R_v}}\times \vec{R}_v)\cdot \hat{z} \nonumber\\ &-2 \lambda \sqrt{\bar{\rho}_s} Q_v Q_s   V(|\vec{R}_v(t)- \vec{R}_s(t)|) \bigg),
\end{align}
%%%%%%%%%%%%%%%%%%%%%%%%%
where we have added a kinetic term of the vortex, which is linear in mass ($M_v$) and quadratic in velocity. Such a term can be generated from the coupling of the superfluid velocity with Cooper pair density $\rho_s$ by considering density fluctuation~\cite{Han:2017fyd}.\\

Similarly, to obtain the action for the collective dynamics of Skyrmion in terms of $R_s(t)$, we write down the part relevant to the Skyrmion motion 
\begin{align}
  S_s&= \int d^3x\, \frac{1}{2\pi} \epsilon_{\mu\nu\rho} \partial^\nu b^\rho \tilde{a}^\mu\nonumber\\ & = \int d^3x\, \left(- a_0 - \frac{m\lambda}{\pi \sqrt{\bar{\rho}_s}} b_i \varepsilon^{ij} \partial_0 a_j\right)
\end{align}
Considering the Skyrmion to be a rigid spin texture i.e. $\hat{n}= \hat{n}(\vec{x}- \vec{R}_s(t))$ and considering spin variation only due to translational motion of the Skyrmion we get the following general equation of motion for Skyrmion (See Appendix~\ref{appendix-E} for details)
%%%%%%%%%%%%%%%%%%%%%%%%%%%%%%%%%%%%%%%%%%%%%%%%%%%%%
%%%%%%%%%%%%%%%%%%%%%%%%%%%%%%%%%%%%%%%%%%%%%%%%%%%%%
%%%%%%%%%%%%%%%%%%%%%%%%%%%%%%%%%%%%%%%%%%%%%%%%%%%%%

%%%%%%%%%%%%%%%%%%%%%%%%%%%%%%%%%%%%%%%%%%%%%%%%%%%%%%%%%%%%%%%%%
%%%%%%%%%%%%%%%%%%%%%%%%%%%%%%%%%%%%%%%%%%%%%%%%%%%%%%%%%%%%%%%%%
%%%%%%%%%%%%%%%%%%%%%%%%%%%%%%%%%%%%%%%%%%%%%%%%%%%%%%%%%%%%%%%%%
\begin{figure}
    \centering
    \includegraphics[width=8cm, height=5.5cm]{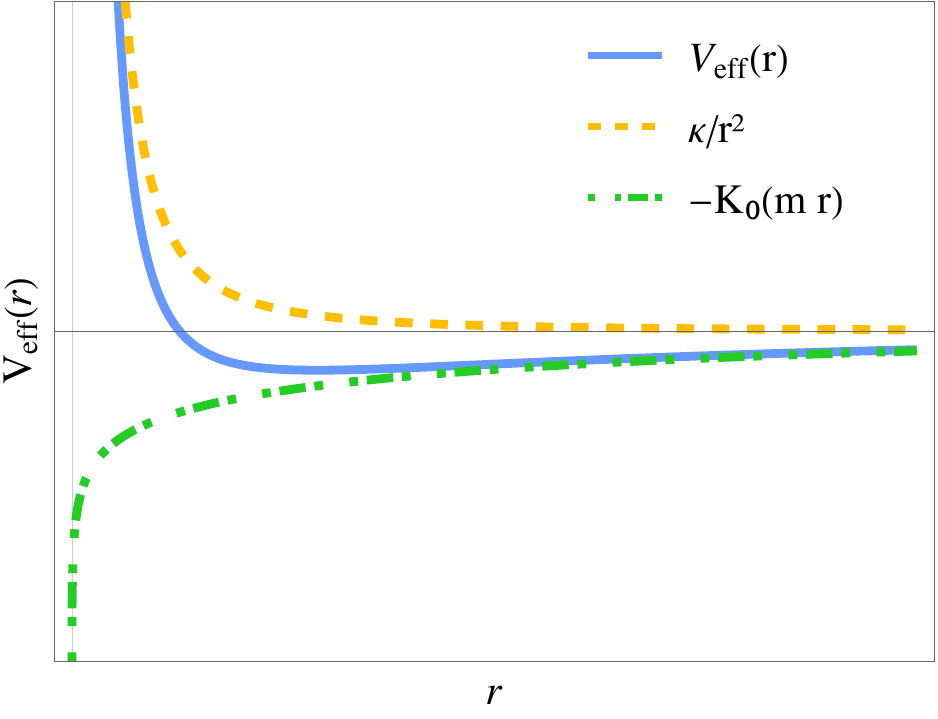} % Image width should be smaller than wrap width
    \caption{Effective potential $V_{\text{eff}}$}
    \label{fig:effective}
\end{figure}
    \begin{align}\label{eom Skyrmion 3}
    S_s = &\int\,dt\, \bigg( \frac{1}{2} M_s \vec{\dot{R}_s}^2 + \frac{Q_s}{4} (1+ m\lambda \sqrt{\bar{\rho}_s})\, \vec{\dot{R_s}}\times \vec{R}_s\nonumber \\ &-  2 \lambda \sqrt{\bar{\rho}_s} Q_v Q_s   V(|\vec{R}_v(t)- \vec{R}_s(t)|) - U_\sigma (\vec{R}_s(t))\bigg).
\end{align}
Here, we have added a mass term for the Skyrmion, considering spin fluctuation in time at the quadratic order. The action for vortex in Eq~\eqref{eom vortex 2} and that for Skyrmion in Eq~\eqref{eom Skyrmion 3} will lead to the following coupled equation of motion
\begin{align}
   & M_v \Ddot{\vec{R}}_v + G_v \hat{z}\times \dot{\vec{R}}_v + \kappa \,\frac{\partial}{\partial_{\vec{R}_v} }V(|\vec{R}_v(t)- \vec{R}_s(t)|)=0.\label{eom vortex 3}\\ &
   M_s \Ddot{\vec{R}}_s + G_s \hat{z}\times \dot{\vec{R}}_s + \kappa \,\frac{\partial}{\partial_{\vec{R}_s} }V(|\vec{R}_v(t)- \vec{R}_s(t)|) + \frac{\partial\, U_\sigma}{\partial_{\vec{R}_s} } =0\label{eom Skyrmion 4}.
\end{align}
where we have written $G_v=\bar{\rho}_s Q_v$, $G_s=\dfrac{Q_s}{2} (1+ m\lambda \sqrt{\bar{\rho}_s})$ which are the effective gyro-coupling for vortex and Skyrmion respectively and $\kappa= 2 \lambda \sqrt{\bar{\rho}_s} Q_v Q_s$ which is the effective coupling constant of the Skyrmion-vortex potential. We notice that due to the interaction between superconducting and ferromagnetic order, the effective gyrocoupling for Skyrmions gets additional contribution--  namely, an interaction induced Magnus force on the Skyrmion. This is another novel contributions of this work. However, detailed study of this novel effect needs a separate volume and we shall keep this discussion for a later work. Also, we can see that the equation of motion of the vortex and the Skyrmion--Eq~\eqref{eom vortex 3} and \eqref{eom Skyrmion 4} are coupled at non-zero values of $\lambda$, which means that the interaction introduced by us in the original action Eq~\eqref{total_lagrangian} leads to coupled dynamics of Skyrmion and vortex. To understand this coupled motion, we consider the following special case.\\

For the following discussion, let us ignore the force due to the non-linear sigma like term and add the Magnus force on the vortex due to superflow $\vec{v}_s$~\cite{feigel1995sign}. This will modify the equation of motion as
\begin{align}\label{coupled eom}
   & M_v \Ddot{\vec{R}}_v + G_v \hat{z}\times \dot{\vec{R}}_v +\kappa \,\frac{\partial V(|\vec{R}_v(t)- \vec{R}_s(t)|)}{\partial_{\vec{R}_v} }+ \hat{z}\times \vec{v}_s=0 .\\ &
   M_s \Ddot{\vec{R}}_s + G_s \hat{z}\times \dot{\vec{R}}_s +\kappa \,\frac{\partial V(|\vec{R}_v(t)- \vec{R}_s(t)|)}{\partial_{\vec{R}_s} } =0,
\end{align}
To proceed we shall consider the centre of mass coordinate $\vec{R}= \dfrac{M_v \vec{R}_v + M_s \vec{R}_s }{M_s + M_v}$ and relative coordinate $\vec{r}= \vec{R}_v - \vec{R}_s$. The equation of motion of the relative coordinate in the limit $G_s<< M_s,\, G_v<< M_v$
\begin{equation}
    \mu \Ddot{\vec{r}}= - \kappa \frac{\partial V(|\vec{r}|)}{\partial \vec{r}} -\frac{\hat{z}\times \vec{v}_s}{M_s},\quad \mu= \frac{M_s M_v}{M_v+M_s}.
\end{equation}
This suggests that the effective potential (See Fig~\ref{fig:effective}) acting on the pair is given by
\begin{equation}
    V_{\text{eff}}= \frac{l^2}{2\mu r^2} -\kappa |V(r)|,
\end{equation}
where the first term arises from the rotational energy of the particle of reduced mass $\mu$ in the reduced one body problem. Also, the negative sign of the second term comes due to the opposite signs of charges of the vortex and Skyrmion configuration considered here. This potential leads to binding of the Skyrmion and the vortex with a finite separation. Now, let us derive the equation of motion of the centre of mass coordinate
\begin{equation}
    \Ddot{\vec{R}}= \hat{z}\times\frac{G_v \dot{\vec{R}}_v + G_s \dot{\vec{R}}_s }{M_s + M_v} - \frac{\hat{z}\times \vec{v}_s}{M_s + M_v}.
\end{equation}
%%%%%%%%%%%%%%%%%%%%%
Considering the limit $\frac{G_v}{M_s + M_v},\,\frac{G_v}{M_s + M_v}<< 1$ we have 
\begin{equation}\label{Skyrmion_Hall}
    \Ddot{\vec{R}}=  - \frac{\hat{z}\times \vec{v}_s}{M_s + M_v}.
\end{equation}

%%%%%%%%%%%%%%%%%%%%%%%%%%
We know that the Magnus force moves the vortex perpendicular to the superflow $\vec{v}_s$~\cite{feigel1995sign}. As the attractive potential leads to a stable pair of anti-Skyrmion and a vortex, with the movement of the vortex perpendicular to the superflow, the centre of mass of the pair also moves in that direction as suggested by the Eq~\eqref{Skyrmion_Hall}. This will create a Hall motion of the Skyrmion due to the drag of vortex. \emph{Thus, we have achieved a novel kind of Skyrmion Hall effect due to the dragging of the vortex.}
\section{Discussion and Outlook:}\label{sec-v}
     In this work, we have developed a field-theoretic description of a ferromagnetic superconductor in which the superconducting and ferromagnetic orders are coupled through a direct interaction term introduced here. This coupling enables the formulation of a dual theory in which topological excitations, namely, the superconducting vortices and magnetic Skyrmions carrying opposite topological charges, interact via an emergent gauge field and can form stable bound pairs. The formation of such composite objects gives rise to several intriguing physical consequences.\\

In particular, when a supercurrent with velocity $\vec{v}_s$ is applied, the vortex experiences a Magnus force, resulting in motion transverse to the direction of the supercurrent. In the present scenario, where the vortex is bound to a Skyrmion, this transverse motion is transmitted to the composite object, leading to a transverse drift of its center of mass. Consequently, the Skyrmion exhibits a Hall-like drift induced by vortex motion. This vortex-drag induced Skyrmion Hall effect is fundamentally distinct from the conventional Skyrmion Hall effect~\cite{PhysRevLett.107.136804, litzius2017Skyrmion} observed in chiral magnets, which arises from spin-transfer torques generated by dissipative electrical currents flowing directly through the Skyrmion texture, rather than from a dissipationless supercurrent.\\

The proposed vortex-drag induced Skyrmion Hall effect can be detected through several experimental probes. One promising platform is hydrazine-treated NbSe$_2$, where superconductivity and ferromagnetism are known to coexist. In such a system, the application of a supercurrent is expected to induce an indirect Hall-like drift of Skyrmions mediated by vortex motion. Since a moving Skyrmion generates an emergent electric field, this drift can be detected electrically, providing a direct signature of the effect. Importantly, because the Skyrmion motion is driven purely by vortex drag, reversing the direction of the applied supercurrent reverses the direction of the Hall-like drift. Therefore, the existence of this effect would be signaled by a change in the sign of the Skyrmion Hall response upon reversal of the supercurrent direction. Moreover, this effect will be suppressed once the superconductivity is suppressed, which can also be used to confirm the existence of this effect. \\

     We also observe that the interaction potential in Eq~\eqref{interaction potential form} arising from the dual model has the following behavior in different distance ranges.
    \begin{align}
        & V(r) \sim -\ln (r/\sqrt{2\pi}\lambda_s) \quad \text{for} \quad r<<\lambda_s\\ &
         V(r) \sim \sqrt{\frac{\pi \lambda_s}{2 r}}\exp\left( (-r/\sqrt{2\pi}\lambda_s)\right) \quad \text{for} \quad r>>\lambda_s
    \end{align}
    where the constant $\lambda_s= 1/\sqrt{2\pi}\tilde{m}$ is the penetration depth of the superconducting order. Thus, within the range $r< \lambda_s$, the attractive interaction is logarithmically diverging, leading to a strongly bound state, while outside this range the attraction fades away. Also, we see that the strength of the interaction depends linearly on $\rho_s$ which means that as the superconducting order becomes weak i.e., $\rho_s \rightarrow 0$, the interaction strength also becomes small. \\

     This leads us to the other very interesting direction, namely, the Berezinskii-Kosterlitz-Thouless (BKT) like physics in the current scenario. As in the BKT scenario, there can two phases- one is characterized by non-zero density of anti-Skyrmion-vortex pairs, while in the other phase the pair density will become zero. If such a phase transition indeed exist that would be a novel scenario in the condensed matter literature. Investigation into this direction is ongoing and will be reported in the future.\\

     We conclude this discussion by briefly commenting on some related works. A closely related dynamical effect was previously studied by Menezes \emph{et al.}~\cite{PhysRevB.100.014431} in the context of ferromagnet--superconductor heterostructures, where it was described as a vortex--Skyrmion Hall effect. Although the phenomenology is related to the vortex-drag-induced Skyrmion Hall effect analyzed in the present work, the two settings differ both in physical platform and in the mechanism responsible for Skyrmion--vortex binding. In Ref.~\cite{PhysRevB.100.014431}, the coupling between the Skyrmion and the vortex is mediated by stray electromagnetic fields in a heterostructure geometry. By contrast, in the present work the binding arises intrinsically from the local interaction term introduced in Eq.~\eqref{proposed interaction}, which directly couples the ferromagnetic and superconducting sectors. This distinction is reflected in the effective vortex--Skyrmion potential derived here, whose interaction strength is controlled explicitly by the coupling constant \(\lambda\).

We also note the related work of Sonin~\cite{PhysRevB.97.224517}, which develops a theoretical description of spin superfluidity and vortex solutions in spin-1 ferromagnetic condensates. While the geometric spin--phase structure appearing in that work provides useful conceptual background, the physical setting considered there is different from that of the present study. In particular, our work concerns an intrinsic ferromagnetic superconductor with distinct superconducting and ferromagnetic sectors, rather than a spinor condensate in which the same degrees of freedom underlie both mass and spin superfluidity. {However, as a coupling between spin and phase, akin to ours, naturally appears within the frame work} of Sonin~\cite{PhysRevB.97.224517}, it may be useful for identifying a possible microscopic origin of the effective interaction proposed here in intrinsic ferromagnetic superconductors. Work along this direction is ongoing and will be reported elsewhere.

%begin{widetext}
\begin{acknowledgments}
    S.M. would like to acknowledge the financial support from the Indian Institute of Technology Bombay through the Institute Post Doctoral Fellowship, and financial support from Anusandhan National Research Foundation (ANRF) through the National Postdoctoral fellowship for the later stages of the work.
\end{acknowledgments}
    \begin{appendix}
    \section{SO$(3)$ invariance of the coupling of the term of Eq~\eqref{proposed interaction}}\label{appendix-A}
  By the definition given in Sec-\ref{sec-ii} of the paper we can write that
\begin{equation}
    a_i = - iz^\dagger \grad z =\frac{1}{2}(1-\cos\theta)\grad \phi.
\end{equation}
With this definition the field strength tensor of this emergent gauge field $a^i$ reads
\begin{equation}
    f_{ij}= \partial_i a_j - \partial_j a_i= \sin\theta (\partial_i\theta \partial_j\phi - \partial_j\theta \partial_i\phi).
\end{equation}

Now, this definition identifies with the expression $\hat{n}\cdot (\partial_i\hat{n}\times \partial_j \hat{n})$ which is related to the Skyrmion density 
\begin{equation}
     \hat{n}\cdot (\partial_i\hat{n}\times \partial_j \hat{n}) = \sin\theta (\partial_i\theta \partial_j\phi - \partial_j\theta \partial_i\phi).
\end{equation}
once we parametrize the spin component as $\hat{n} = (\sin\theta\cos\phi, \sin\theta\sin\phi, \cos\theta )$. This connection is also mentioned in Eq.10 and Eq. 11. It can be seen that the expression $\hat{n}\cdot (\partial_i\hat{n}\times \partial_j \hat{n})$ is invariant under global SO($3$) transformation of the spin $\hat{n}$ as follows:\\

Under global SO($3$) components of $\hat{n}$ transforms as $n_\alpha \rightarrow R_{\alpha\beta} n_\beta$, where $R$ matrices represent SO($3$) transformation matrices. With this $\hat{n}\cdot (\partial_i\hat{n}\times \partial_j \hat{n})$ transforms as
\begin{equation}
    \hat{n}\cdot (\partial_i\hat{n}\times \partial_j \hat{n})\rightarrow \varepsilon^{\alpha\beta\gamma} R^{\alpha a} R^{\beta b} R^{\gamma c} n^a \partial_i n^b \partial_j n^c,
\end{equation}
where $(a, b, c, \alpha, \beta, \gamma) \in (1, 2, 3)$ . Using the identity $\varepsilon^{\alpha\beta\gamma} R^{\alpha a} R^{\beta b} R^{\gamma c}= \det(R) \varepsilon^{abc}$ and putting $\det(R)= 1$ we get

\begin{equation}
    \hat{n}\cdot (\partial_i\hat{n}\times \partial_j \hat{n})\rightarrow \varepsilon^{abc}  n^a \partial_i n^b \partial_j n^c=  \hat{n}\cdot (\partial_i\hat{n}\times \partial_j \hat{n}).
\end{equation}
Through this derivation we saw that $f_{ij}= \hat{n}\cdot (\partial_i\hat{n}\times \partial_j \hat{n})$ remains under global SO($3$) transformation of the unit spin field. This immediately implies that under this global SO(3) transformation the $\mathbb{C}P^1$ gauge field $a_i$ changes by a local gauge transformation: $a_i \rightarrow a_i + \partial_i \Lambda_R$ and thus the interaction term $\mathcal{L}_I$ changes as 
\begin{equation}
    \delta\mathcal{L}_I = \partial_i \Lambda_R (\mathbf{\grad}\chi + i\eta_v^*\mathbf{\grad}\eta_v + q\vec{A}).
\end{equation}
Recognizing $(\mathbf{\grad}\chi + i\eta_v^*\mathbf{\grad}\eta_v + q\vec{A})$ to be proportional to conserved current of the superconducting component $\vec{J}_s$, we see that the above term contributes only a total derivative term in the action i.e.
\begin{equation}
    \delta S_I \propto  \int d^3x\, \grad \cdot (\Lambda_R \vec{J}_s),\, \grad \cdot (\vec{J}_s)=0
\end{equation}
This term becomes zero for a system with a boundary at spatial infinity as the current term decays to zero. Thus, although the density in Eq. (17) is not strictly SO(3)-invariant, the corresponding action and dynamics are SO(3)-invariant (upto a boundary term).
%\vspace{0.2 cm}
 \section{Dualization of $\mathcal{L}_T$}\label{appendix-B}
    We start the derivation of the dual theory from the following effective theory of a ferromagnetic superconductors. 
    \small{\begin{align}\label{spinon rep app}
& \mathcal{L}_T\nonumber \\ & = \rho_s(\partial_t\chi + i \eta_v^*\partial_t \eta_v + q A_0 - \frac{1}{\rho_s} a_0)
- \frac{\bar{\rho}_s}{2m}\left(\mathbf{\grad}\chi + i\eta_v^*\mathbf{\grad}\eta_v + q\vec{A}\right)^2 \nonumber \\ & - \frac{g}{2} \left(\rho_s - \bar{\rho}_s\right)^2  - \frac{\rho_M}{2} 4(|D_i z|^2) - \lambda \sqrt{\bar{\rho}_s}\vec{a} \cdot (\mathbf{\grad}\chi + i\eta_v^*\mathbf{\grad}\eta_v + q\vec{A}),
\end{align}}

To proceed towards the formulation of the duality, we first combine the second and the last term in Eq~\eqref{spinon rep app} by the completion of square as follows
\begin{align}\label{completion_of_square app}
  & - \frac{\bar{\rho}_s}{2m}(\mathbf{\grad}\chi + i\eta_v^*\mathbf{\grad}\eta_v + q\vec{A})^2 - \lambda \sqrt{\bar{\rho}_s} \mathbf{a} \cdot \left(\mathbf{\grad}\chi + i\eta_v^*\mathbf{\grad}\eta_v + q\vec{A}\right)\nonumber\\ &= -\frac{\bar{\rho}_s}{2m}\left(\mathbf{\grad}\chi + i\eta_v^*\mathbf{\grad}\eta_v + q\vec{A} + \frac{2m\lambda}{\sqrt{\bar{\rho}_s}} \vec{a}\right)^2 +2m\lambda^2 |\mathbf{a}|^2.
\end{align}
This leads to the following form of the Lagrangian
%\begin{widetext}
\begin{align}\label{cp1 rep1 app}
    &\mathcal{L}= \rho_s\left(\partial_t\chi + i \eta_v^*\partial_t \eta_v + q A_0 - \frac{1}{\rho_s} a_0\right)\nonumber\\ & -\frac{\bar{\rho}_s}{2m}\left(\mathbf{\grad}\chi + i\eta_v^*\mathbf{\grad}\eta_v + q\vec{A} + \frac{2m\lambda}{\sqrt{\bar{\rho}_s}} \vec{a}\right)^2\nonumber \\ &- \frac{g}{2} \left(\rho_s - \bar{\rho}_s\right)^2  -  4\left(\frac{\rho_M}{2}|\mathbf{\grad} z|^2 - \left(\frac{\rho_M}{2} +\frac{m\lambda^2}{2}\right)|\mathbf{a}|^2\right).
\end{align}
   %\end{widetext}
   
The last term in the Eq~\eqref{cp1 rep1 app} can be written in a much nicer form, which will let us understand the meaning of the extra term added and subtracted in Eq~\eqref{completion_of_square app}. First, we rewrite the spinor field $z= \zeta w$, where $\zeta$ is a constant to be determined later. Due to this redefinition, the emergent gauge field $\vec{a}$ changes as
\begin{equation}
    \vec{a}= -i z^\dagger \grad z = -i \zeta^2 w^\dagger \mathbf{\grad} w = \zeta^2 \vec{a}^\prime.
\end{equation}
This will change the concerned term in the following way
\begin{align}
    &\left(\frac{\rho_M}{2}|\mathbf{\grad} z|^2 - \left(\frac{\rho_M}{2} +\frac{m\lambda^2}{2}\right)|\mathbf{a}|^2\right)\nonumber\\ & =  \frac{\rho_M}{2} \zeta^2 (|\mathbf{\grad} w|^2 - \zeta^2\left(1 +\frac{m\lambda^2}{\rho_M}\right)|\vec{a}^\prime|^2).
\end{align}
If we chose the constant $\zeta^2= \dfrac{1}{\left(1 +\frac{m\lambda^2}{\rho_M}\right)}$ then the term takes the form of a $\mathbb{C}\text{P}^1$ representation of the non-linear sigma term:
\begin{align}
    &\left(\frac{\rho_M}{2}|\mathbf{\grad} z|^2 - \left(\frac{\rho_M}{2} +\frac{m\lambda^2}{2}\right)|\mathbf{a}|^2\right) =  \frac{\tilde{\rho}_M}{2} (|\mathbf{\grad} w|^2 - |\vec{a}^\prime|^2),\\ & \text{where}\quad\frac{\tilde{\rho}_M}{2} =\frac{\rho_M}{\left(2 +\frac{2m\lambda^2}{\rho_M}\right)}.
\end{align}
Thus, we see that the concerned term resembles the non-linear sigma term in $\mathbb{C}\text{P}^1$ representation with renormalized spin-stiffness. Taking into account all these changes, we write down the total Lagrangian that will be dualized as follows

\begin{align}\label{Lagrangian befor dualization}
  &  \mathcal{L}=   \rho_s\left(\partial_t\chi + i \eta_v^*\partial_t \eta_v + q A_0 - \frac{1}{\rho_s} a_0\right)\nonumber \\ & -\frac{\bar{\rho}_s}{2m}\left(\mathbf{\grad}\chi + i\eta_v^*\mathbf{\grad}\eta_v + q\vec{A} + \frac{2m\lambda}{\sqrt{\bar{\rho}_s}} \vec{a}\right)^2\nonumber \\ &- \frac{g}{2} \left(\rho_s - \bar{\rho}_s\right)^2  -  2 \tilde{\rho}_M (|\mathbf{\grad} w|^2 - |\vec{a}^\prime|^2)
\end{align}

To derive the dual model, let us first linearize the first term in the above Lagrangian by introducing an auxiliary vector field $f_i$ through a Hubbard Stratonovich (H-S) transformation as follows
\begin{align}
   & \int \mathcal{D}\vec{f}  \exp \bigg[ i\int d^3x  \nonumber \\ & \left(\vec{f} -  \sqrt{\frac{\bar{\rho}_s}{2m}}\left(\mathbf{\grad}\chi + i\eta_v^*\mathbf{\grad}\eta_v + q\vec{A} + \frac{2m\lambda}{\sqrt{\bar{\rho}_s}} \vec{a}\right)\right)^2\bigg]\nonumber \\ & = \text{ A constant}
\end{align}

%%%%%%%%%%
Thus, the total Lagrangian thus obtained is given by
\begin{align}
   & \mathcal{L}= \rho_s\left(\partial_t\chi + i \eta_v^*\partial_t \eta_v + q A_0 - \frac{1}{\rho_s} a_0\right) \nonumber\\ &+ |\vec{f}|^2 -\sqrt{\frac{2\bar{\rho}_s}{m}} \vec{f}\cdot \left(\mathbf{\grad}\chi + i\eta_v^*\mathbf{\grad}\eta_v + q\vec{A} + \frac{2m\lambda}{\sqrt{\bar{\rho}_s}} \vec{a}\right)\nonumber \\ & - \frac{g}{2} \left(\rho_s - \bar{\rho}_s\right)^2  -  2 \tilde{\rho}_M (|\mathbf{\grad} w|^2 - |\vec{a}^\prime|^2).
\end{align}

This linearized Lagrangian can be put into a more concise form by first absorbing the factor $\sqrt{\frac{2\rho_s}{m}}$ into a redefinition of the auxiliary field  and $\mathbb{C}\text{P}^1$ gauge field 
\begin{align}
   J_0=\rho_s,\quad J_i= \sqrt{\frac{2\bar{\rho}_s}{m}} f_i,\quad \tilde{a}_0= -\frac{1}{\rho_s} a_0,\quad \tilde{\vec{a}}= \frac{2m\lambda}{\sqrt{\bar{\rho}_s}} \vec{a}
\end{align}
This redefinition allows us to write the first two terms using relativistic notation and lead us to the following Lagrangian
\begin{align} \label{auxiliary lagrangian}
   \mathcal{L}= &\frac{m}{2\bar{\rho}_s}J^i J^i + J^\mu \left(\partial_\mu\chi + i \eta_v^*\partial_\mu \eta_v + q A_\mu + \tilde{a}_\mu\right)\nonumber \\ &- \frac{g}{2} \left(J_0 - \bar{\rho}_s\right)^2  -  2 \tilde{\rho}_M (|\mathbf{\grad} w|^2 - |\vec{a}^\prime|^2).,
\end{align}
We have previously divided the phase of the superconducting order parameter into two parts, $\chi$ and $\chi_v$, among which the first part  is the small scale fluctuation, while the second part is the singular or multi-valued part, which incorporates the large scale topological excitations like vortices~\cite{Mukherjee:2019vmi}. To obtain the theory of vortices and Skyrmions, one can integrate these small scale fluctuations $\chi$. To proceed, one can take an integration by parts and obtain from the above Lagrangian
\begin{align}
    \mathcal{L}= &\frac{m}{2\bar{\rho}_s}J^i J^i -\chi \partial_\mu J^\mu +J^\mu \left( i \eta_v^*\partial_\mu \eta_v + q A_\mu + \tilde{a}_\mu\right)\nonumber\\ & - \frac{g}{2} \left(J_0 - \bar{\rho}_s\right)^2  -  2 \tilde{\rho}_M (|\mathbf{\grad} w|^2 - |\vec{a}^\prime|^2).
\end{align}
Then functional integral over $\chi$ gives us 
\begin{equation}
    \int \mathcal{D}\chi\, e^{- i \,\int\, d^3x\, \left(\partial_\mu J^\mu\right)}= \delta\left(\partial_\mu J^\mu\right)
\end{equation}
This delta function condition can be solved in $2+1\,d$ by the assumption of a new emergent vector field $b^\mu$ as follows
 \begin{equation}
     \partial^\mu J_\mu=0 \implies J_\mu= \frac{1}{2\pi} \epsilon_{\mu\nu\rho} \partial^\nu b^\rho,
 \end{equation}
 where $b^\mu$ is an emergent gauge field. This will lead to the following form of the dual Lagrangian
 \begin{align}\label{dual_lagrangian_1 appendix}
   \mathcal{L}= &\frac{m}{8\pi^2\bar{\rho}_s} (\partial_0 b_i-\partial_i b_0 )^2  - \frac{g}{8\pi^2} \left(\grad \times \vec{b} - 2\pi \bar{\rho}_s\right)^2 +\nonumber\\ & \frac{1}{2\pi} \epsilon_{\mu\nu\rho} \partial^\nu b^\rho\left( i \eta_v^*\partial^\mu \eta_v + q A^\mu + \tilde{a}^\mu\right) - 2\tilde{\rho}_M (|\mathbf{\grad} w|^2 - |\vec{a}^\prime|^2).
   \end{align}

   \section{Integrating out $A_\mu$:}\label{appendix-C}
The generating functional for the electromagnetic vector potential $A_\mu$ is given by
%%%%%%%%%%%%%%%%%%%%%%%%%%
\begin{equation}
    \mathcal{Z}_A= \int \mathcal{D}A_\mu \exp{i\int\, d^3x \, \left(-\frac{1}{4}F^{\mu\nu}F_{\mu\nu} +\frac{q}{2\pi} A_\mu \varepsilon^{\mu\nu\lambda} \partial^\nu b^\lambda\right) }
\end{equation}
One can rearrange the action in the above written generating functional as follows
\begin{align}
    S_A = &\int d^3x\, \left( A_\mu M^{\mu\nu} A_\nu +\frac{q}{2\pi} A_\mu \varepsilon^{\mu\nu\lambda} \partial^\nu b^\lambda\right),\\ & M^{\mu\nu}= g^{\mu\nu} \partial^2 - (1-\frac{1}{\xi}) \partial^\mu \partial^\nu.
\end{align}
Here $M^{\mu\nu}$ is the inverse propagator, and the extra contribution containing $1/\xi$ comes due to gauge fixing. One can now use the photon propagator $G^{\mu\nu}$, which is the inverse of $M^{\mu\nu}$, to integrate out $A_\mu$ by first completing the square and decoupling $A_\mu$ from $b_\mu$. These steps will result in the following generating functional
%\begin{widetext}
\begin{align}
   & \mathcal{Z}_A \nonumber = \int \mathcal{D} A^\prime_\mu \\ &  \exp \bigg[i\int\, d^3x \, \bigg[A^\prime_\mu M^{\mu\nu} A^\prime_\nu  - \frac{q^2}{16\pi^2} (\varepsilon^{\mu\nu\lambda}\partial_\nu b_\lambda) (\square)^{-1} (\varepsilon_{\mu\alpha\beta}\partial^\alpha b^\beta)\bigg]\bigg]
\end{align}
%\end{widetext}
Here we have written 
\begin{equation}
    A^\prime_\mu = A_\mu + \frac{q}{4\pi} \int d^4y\, G_{\mu\nu}(x-y)  (\varepsilon^{\nu\alpha\beta}\partial_\alpha b_\beta)
\end{equation} 
Thus, after integrating out the vector field $A_\mu$ we get the following new dual Lagrangian 
%\begin{widetext}
\begin{align}
   & \mathcal{L}= \frac{m}{8\pi^2\bar{\rho}_s} (\partial_0 b_i-\partial_i b_0 )^2  - \frac{g}{8\pi^2} \left(\grad \times \vec{b} - 2\pi \bar{\rho}_s\right)^2  \nonumber \\ & +  \frac{1}{2\pi} \epsilon_{\mu\nu\rho} \partial^\nu b^\rho\left( i \eta_v^*\partial^\mu \eta_v + \tilde{a}^\mu\right)\nonumber\\ & - \frac{q^2}{16\pi^2} (\varepsilon^{\mu\nu\lambda}\partial_\nu b_\lambda) (\square)^{-1} (\varepsilon_{\mu\alpha\beta}\partial^\alpha b^\beta)  - 2 \tilde{\rho}_M (|\mathbf{\grad} w|^2 - |\vec{a}^\prime|^2).
\end{align}    
%\end{widetext}

\section{Integrating out $b_0$}\label{appendix-D}

\begin{widetext}
 To proceed we shall separate out the terms with $b_0$ and write down the corresponding Lagrangian
    \begin{equation}
    \mathcal{Z}_{b_0}= \int \mathcal{D}b_0 \exp{i \int d^3x\, \left[-\frac{m}{8\pi^2 \bar{\rho}_s}\,b_0 \left( \grad^2 - \frac{q^2\bar{\rho}_s}{2m}\right) b_0 + \frac{b_0}{2\pi} \left(J^0_v + \frac{2m\lambda}{\sqrt{\bar{\rho}_s}} J^0_s \right)\right] }
\end{equation}
Now integrating out the field $b_0$ we get
\begin{equation}
    \mathcal{Z}_{b_0}= \int \mathcal{D}b^\prime_0 \exp{-i \int d^3x\, \left( b^\prime_0 \Delta^{-1} b^\prime_0 \right)} \exp{\frac{i}{4}\int\, d^3x\,d^3y\, \left( \frac{1}{4\pi^2} \left(J^0_v + \frac{2m\lambda}{\sqrt{\bar{\rho}_s}} J^0_s \right) \Delta(x-y) \left(J^0_v + \frac{2m\lambda}{\sqrt{\bar{\rho}_s}} J^0_s \right) \right)}
\end{equation}
\end{widetext}

Here $\Delta(x-y)$ is the Green's function for the operator $\frac{m}{8\pi^2 \bar{\rho}_s}\, \left( \grad^2 - \frac{q^2\bar{\rho}_s}{2m}\right)$. Below we shall consider the situation where a vortex and a Skyrmion is present and interact. In such a case only the Skyrmion-vortex interaction term will be non-zero. Putting the Skyrmion and vortex charge density $J_s^0$ and $J_v^0$ as
\begin{equation}
    J_s^0= Q_s \delta^2(\vec{x}- \vec{R}_s),\quad J_v^0= Q_v \delta^2(\vec{x}- \vec{R}_v).
\end{equation}
\begin{widetext}
we get 
    \begin{equation}\label{interaction potential appendix}
    \mathcal{Z}_{b_0}= \int \mathcal{D}b^\prime_0 \exp{-i \int d^3x\, \left( b^\prime_0 \Delta^{-1} b^\prime_0 \right)} \exp{-i\int\, dt\, \lambda\sqrt{\bar{\rho}_s} Q_v Q_s   V(|\vec{R}_v(t)- \vec{R}_s(t)|) },
\end{equation}
where 
\begin{equation}
    V(|\vec{R}_v- \vec{R}_s|)=  \int \,\frac{d^2k}{(2\pi)^2} \, \frac{e^{i\vec{k}\cdot (\vec{R}_v- \vec{R}_s)}}{k^2 + \frac{q^2\bar{\rho}_s}{2m} }= \frac{1}{2\pi} K_0\left[\sqrt{\frac{q^2\rho_s}{2m}} |\vec{R}_v- \vec{R}_s|\right]
\end{equation}
\end{widetext}

\section{Equation of motion of a Skyrmion }\label{appendix-E}
Below, we shall consider the variation of this action due to the Skyrmion motion. This will give us
\setlength{\parskip}{-0.2em}
\begin{equation}\label{eom of Skyrmion 1}
  \delta S_s= \int d^3x\, \left(- \delta a_0 - \frac{m\lambda}{\pi \sqrt{\bar{\rho}_s}} b_i \varepsilon^{ij} \partial_0 \delta a_j\right)
\end{equation}
Now one can show that 
\begin{align}
    &\delta a_0 = -(\hat{n}\times \partial_t \hat{n})\cdot \delta \hat{n} = \frac{1}{2}\sin\theta \delta\theta \dot{\phi} + \frac{1}{2}(1-\cos\theta)\delta\dot{\phi}\\ & 
    \delta a_i = -(\hat{n}\times \partial_i \hat{n})\cdot \delta \hat{n} = \frac{1}{2}\sin\theta \delta\theta \partial_i\phi + \frac{1}{2}(1-\cos\theta)\partial_i \delta\phi
\end{align}
These expressions, when plugged into Eq~\eqref{eom of Skyrmion 1} we have
\begin{equation}\label{eom of Skyrmion 2}
    S_s= \int\, d^3x\, \frac{1}{2}\left((\hat{n}\times \partial_t \hat{n})\cdot \delta \hat{n} - \frac{m\lambda}{\pi \sqrt{\bar{\rho}_s}} \varepsilon^{ij}\partial_0 b_i  (\hat{n}\times \partial_j \hat{n})\cdot \delta \hat{n}\right).
\end{equation}
As stated previously here we shall consider that variation of spin field is generated only through Skyrmion motion. This can be implemented by the assumption that $\hat{n}= \hat{n}(\vec{x}- \vec{R}_s)$. This will give us the following
\begin{equation}
    \delta\hat{n}= -\frac{\partial\hat{n}}{\partial x} \delta X_s -\frac{\partial\hat{n}}{\partial y} \delta Y_s
\end{equation}
Using this and the expression for $\vec{b}$ into Eq~\eqref{eom of Skyrmion 2} we get
    \begin{align}
    \delta S_s= &\int\, d^3 x\, \frac{1}{2} \bigg( Q_s \delta^2(\vec{x}-\vec{R}_s) (\dot{X}_s \delta Y_s - \dot{Y}_s \delta X_s )\nonumber\\ & + m\lambda \sqrt{\bar{\rho}_s} Q_s \delta^2(\vec{x}-\vec{R}_s) (\dot{X}_s \delta Y_s - \dot{Y}_s \delta X_s ) \bigg),
\end{align}
where we have written $\hat{n}\cdot (\partial_x \hat{n}\times \partial_y \hat{n}) = Q_s \delta^2 (\vec{x}-\vec{R}_s)$, the Skyrmion charge. Now using the fact that
\begin{equation}
    (\dot{X}_s \delta Y_s - \dot{Y}_s \delta X_s ) = \frac{1}{2} \delta (\dot{X}_s\, Y_s - \dot{Y}_s \, X_s ) = \frac{1}{2} \delta (\vec{\dot{R_s}}\times \vec{R}_s)
\end{equation}
we can write the Skyrmion action as
\begin{equation}
    S_s = \int\,dt\, \frac{Q_s}{4} (1+ m\lambda \sqrt{\bar{\rho}_s})\, \vec{\dot{R_s}}\times \vec{R}_s.
\end{equation}
In presence of the Skyrmion-vortex interaction and taking into account the effect of the modified non-linear sigma term the above action becomes
\begin{widetext}
    \begin{equation}
    S_s = \int\,dt\, \left( \frac{Q_s}{4} (1+ m\lambda \sqrt{\bar{\rho}_s})\, \vec{\dot{R_s}}\times \vec{R}_s - 2 \lambda \sqrt{\bar{\rho}_s} Q_v Q_s   V(|\vec{R}_v(t)- \vec{R}_s(t)|) - U_\sigma (\vec{R}_s(t))\right),
\end{equation}
where we have written
\begin{equation}
    U_\sigma (\vec{R}_s(t))= \int \,d^2x\, 4 \tilde{\rho}_M (|\mathbf{\grad} w|^2 - |\vec{a}^\prime|^2).
\end{equation}
\end{widetext}
\end{appendix}
%\end{widetext}

\bibliography{bound}
\end{document}